\documentclass[useAMS,usenatbib]{mn2e}
\usepackage{amssymb}
\usepackage{natbib}
\usepackage{times}
\usepackage{graphics,epsfig}
\usepackage{graphicx}
\usepackage{amsmath}
\usepackage{comment}
\usepackage{supertabular}

\usepackage{graphicx}
\usepackage{amsmath}
\usepackage{color}

\newcommand{\kms}{km~s$^{-1}$}

\newcommand{\hi}{H{\sc i}}
\newcommand{\cm}{cm$^{-2}$}
\newcommand{\nhi}{\ensuremath{\rm N_{HI}}}
\newcommand{\hii}{H{\sc i}~21cm}

\newcommand{\ts}{\ensuremath{\rm T_{s}}}
\newcommand{\tk}{\ensuremath{\rm T_{k}}}
\newcommand{\td}{\ensuremath{\rm T_{D}}}

\title {Detection of the Galactic Warm Neutral Medium in H{\sc i} 21cm absorption}
\author [Patra et al.]{	Narendra Nath Patra$^{1}$\thanks{E-mail: narendra@ncra.tifr.res.in}, Nissim Kanekar$^1$\thanks{DST Swarnajayanti Fellow}, Jayaram N. Chengalur$^{1}$ and Nirupam Roy$^2$\\
$^{1}$National Centre for Radio Astrophysics, Tata Institute of Fundamental Research, Pune University campus, Pune 411 007, India \\	
$^{2}$Indian Institute of Science, Bengaluru, India 
}
\date {}
\begin {document}
\maketitle
%\pagerange{\pageref{firstpage}--\pageref{lastpage}} \pubyear{}
%\label{firstpage}

\begin{abstract}
We report a deep Giant Metrewave Radio Telescope (GMRT) search for Galactic H{\sc i} 21cm 
absorption towards the quasar B0438$-$436, yielding the detection of wide, weak H{\sc i} 
21cm absorption, with a velocity-integrated H{\sc i} 
21cm optical depth of $0.0188 \pm 0.0036$~km~s$^{-1}$. Comparing this with the H{\sc i} 
column density measured in the Parkes Galactic All-Sky Survey gives a column density-weighted 
harmonic mean spin temperature of $3760 \pm 365$~K, one of the highest measured in the 
Galaxy. This is consistent with most of the H{\sc i} along the sightline arising in the stable 
warm neutral medium (WNM). The low peak H{\sc i} 21cm optical depth towards B0438$-$436 implies 
negligible self-absorption, allowing a multi-Gaussian joint decomposition of the H{\sc i} 21cm 
absorption and emission spectra. This yields a gas kinetic temperature of $\rm T_k \leq 
(4910 \pm 1900)$~K, and a spin temperature of $\rm T_s = (1000 \pm 345)$~K for the gas that 
gives rise to the H{\sc i} 21cm absorption. Our data are consistent with the 
H{\sc i} 21cm absorption arising from either the stable WNM, with $\rm T_s \ll T_k$, $\rm T_k \approx 5000$~K,
and little penetration of the background Lyman-$\alpha$ radiation field into the neutral hydrogen,
or from the unstable neutral medium, with $\rm T_s \approx T_k \approx 1000\;K$.
\end{abstract}


\section{Introduction}

Neutral hydrogen (\hi) is a key constituent of the interstellar medium (ISM) of galaxies, making 
up almost half the ISM in the Milky Way. The earliest \hii\ studies \citep[e.g.][]{clark65} found 
evidence that \hi\ exists in two phases, the cold neutral medium (CNM), which gives rise to the 
narrow, deep components seen in \hii\ absorption spectra towards background radio 
sources, and the warm neutral medium (WNM), which contributes to the relatively smooth and broad 
\hii\ emission profiles but has very weak \hii\ absorption. Similarly, theoretical models indicate 
that Galactic \hi\ exists in one of the above two stable phases, with low kinetic temperatures 
($\tk \approx 40-300$~K) and high number densities ($\approx 10-100~{\rm cm}^{-3}$) in the CNM, and high 
temperatures ($\tk \approx 5000 - 8000$~K) and low number densities ($\approx 0.1-1~{\rm cm}^{-3}$)
 in the WNM \citep[e.g.][]{field69,mckee77,wolfire95,wolfire03}. \hi\ at intermediate temperatures 
($\approx 500-5000$~K) is expected to be unstable, and to quickly move into one of the two 
stable phases. 

Over the last few decades, \hii\ absorption studies have made much progress in characterising conditions 
in the CNM in the Galaxy
\citep[e.g.][]{rad72,dickey78,payne83,heiles03a,roy13a}, finding that the CNM indeed has $\tk \approx 20-300$~K, 
consistent with theoretical expectations. However, the complexity of the \hii\ emission 
profiles and the difficulty of detecting the WNM in \hii\ absorption have made it difficult 
to verify the prediction of a stable warm phase with $\tk \approx 5000-8000$~K. Indeed, 
attempts at measuring the WNM temperature by fitting multi-Gaussian models to either a combination 
of \hii\ emission and absorption spectra \citep{heiles03a,heiles03b}, or interferometric \hii\ 
absorption spectra \citep{kanekar03b,roy13b,murray15} have found significant fractions of the 
\hi\ to be in the thermally unstable phase, with $\tk \approx 1000$~K. Very few of the 
above Gaussian components show temperatures within or larger than the stable WNM range, 
$\approx 5000-8000$~K. However, most of the sightlines in the above studies have complex 
\hii\ absorption profiles, implying that one is typically searching for WNM absorption 
in the presence of far stronger, multi-component CNM absorption.
 
The best sightlines for a reliable detection of the WNM in \hii\ absorption, and for 
an accurate estimate of the WNM kinetic temperature, are those with the least complexity 
in the \hii\ absorption profile. Our earlier interferometric 
\hii\ absorption survey of compact radio sources achieved root-mean-square (RMS) \hii\ optical 
depth noise values of $\approx 0.001$ per $\approx 1$~\kms\ channel, and yielded detections of 
\hii\ absorption in 33 of 34 sightlines.  The sole sightline without a detection of absorption, 
towards the quasar B0438$-$436, has one of the lowest \hi\ column densities of the sample 
\citep[$\nhi = 1.29 \times 10^{20}$~\cm;][]{kalberla15}, and is at a high Galactic latitude, implying 
low CNM contamination. The absence of CNM absorption on this sightline, despite the high sensitivity 
of our survey, suggests that B0438$-$436 is a good target for a clean search for the WNM in 
absorption. We hence used the Giant Metrewave Radio Telescope (GMRT) to carry out a deep 
search for \hii\ absorption towards B0438$-$436, the results of which are 
described in this {\it Letter}.

\section{Observations, data analysis, and results}
\label{sec:obs}

% Our original GMRT observations of B0438$-$436, described in \citet{roy13a}, used a 0.5~MHz 
% bandwidth (sub-divided into 256 channels), the GMRT hardware correlator, and a total observing 
% time of $\approx 10$~hours. A significant fraction of these data were lost due to problems with 
% the hardware correlator, and the narrow observing bandwidth meant a total usable velocity coverage 
% of only $\approx 90$~\kms. 
Our GMRT search for Galactic \hii\ absorption towards B0438$-$436 was carried out over 
May~$6-15$, 2011, using the L-band receivers, with a total observing time of 
$\approx 30$~hours over 6 observing sessions. The GMRT Software Backend (GSB) was used as the 
correlator, with a bandwidth of $\approx 1.067$~MHz sub-divided into 512~channels, yielding 
a total velocity coverage of $\approx 225$~\kms\ and a velocity resolution of $\approx 0.43$~\kms. 
We used frequency-switching at the first GMRT local oscillator, on B0438$-$436 itself,
to calibrate the system passband; the switching was carried out every five minutes with a 
throw of 5~MHz. Observations of 3C48 and 3C147 at the start and end of each run were used 
to calibrate the flux density scale. Since B0438$-$436 is a phase 
calibrator for the GMRT, no additional phase calibration was necessary. The use of the GSB 
for these observations implied a far superior data quality and a larger total velocity coverage 
than those of our earlier GMRT observations of B0438$-$436 \citep{roy13a}. Allied with our higher 
sensitivity, this significantly improved our ability to detect wide \hii\ absorption lines.

All data were analysed in the Astronomical Image Processing System \citep[AIPS;][]{greisen03}, 
following standard data editing, calibration, self-calibration, imaging, and continuum 
subtraction procedures \citep[e.g.][]{roy13a}. The data from the observing 
run on 6~May were found to be severely affected by RFI, and were hence excluded from the later 
analysis. The task {\sc cvel} was finally used to shift the residual visibilities to the 
local standard of rest (LSR) velocity frame. These visibilities were then imaged to produce 
a spectral cube, using natural weighting and excluding baselines shorter than $1$~k$\lambda$ 
to reduce contamination from \hii\ emission within the primary beam. The \hii\ spectrum was 
obtained by taking a cut through the spectral cube at the location of the quasar; a second-order
baseline was then fitted to line-free regions, and subtracted out, to obtain the final 
\hii\ absorption spectrum. 

We measured a flux density of 4.2~Jy for B0438$-$436 from the final GMRT continuum image; the 
error on this value is dominated by uncertainties in the GMRT flux density scale, which we 
estimate to be $\approx 15$\%. The flux density is in reasonable agreement with the 
value of 5.0~Jy listed in the Very Large Array Calibrator Manual.

For Galactic \hii\ absorption studies, the spectral RMS noise depends on the 
observing frequency, due to the contribution from the brightness temperature of the \hi\ 
emission in the beam \citep{roy13a}. We followed the procedure of \citet{roy13a} to combine the 
brightness temperature measured in the Parkes Galactic All-Sky Sky Survey 
\citep[GASS;][]{mccluregriffiths09,kalberla15} \hii\ emission spectrum with the GMRT 
system temperature ($\approx 73$~K at 1420~MHz), to determine the RMS noise spectrum. Note that 
this correction is very small for B0438$-$436 as the peak \hii\ brightness temperature in 
the GASS spectrum is only $\approx 2$~K, far lower than the GMRT system temperature 
at the observing frequency. The final optical depth RMS noise on the GMRT spectrum is 
$\approx 1.0 \times 10^{-3}$ per $0.43$~\kms\ channel. 

Fig.~\ref{fig:spc}[A] shows our final Galactic \hii\ absorption spectrum towards B0438$-$436, with 
\hii\ optical depth plotted versus LSR velocity, after Hanning-smoothing and re-sampling the spectrum, 
at a velocity resolution of $\approx 0.86$~\kms\ (with an optical depth RMS noise of $\approx 0.72 
\times 10^{-3}$~per 0.86~\kms\ channel). Weak absorption can be seen close to zero LSR velocity. 
Fig.~\ref{fig:spc}[B] shows the spectrum after additional smoothing and re-sampling, at a velocity 
resolution of $\approx 6.0$~\kms; the RMS optical depth noise on this spectrum is 
$\approx 2.4 \times 10^{-4}$ per $\approx 6.0$~\kms. The \hii\ absorption is now clearly visible, 
detected at $\approx 5.3\sigma$ significance: the velocity-integrated \hii\ optical depth 
(over line channels) is $\int \tau d{\rm V} = (0.0188 \pm 0.0036)$~\kms. The rest of the 
spectrum, away from the absorption feature, shows no evidence for any structure in the baseline.
The Kolmogorov-Smirnov rank-1 and Anderson-Darling tests find that the off-line channels are 
consistent with being drawn from a Gaussian distribution.

\section{Discussion}

\begin{figure}
\begin{center}
\begin{tabular}{c}
	\includegraphics[width=3.3in]{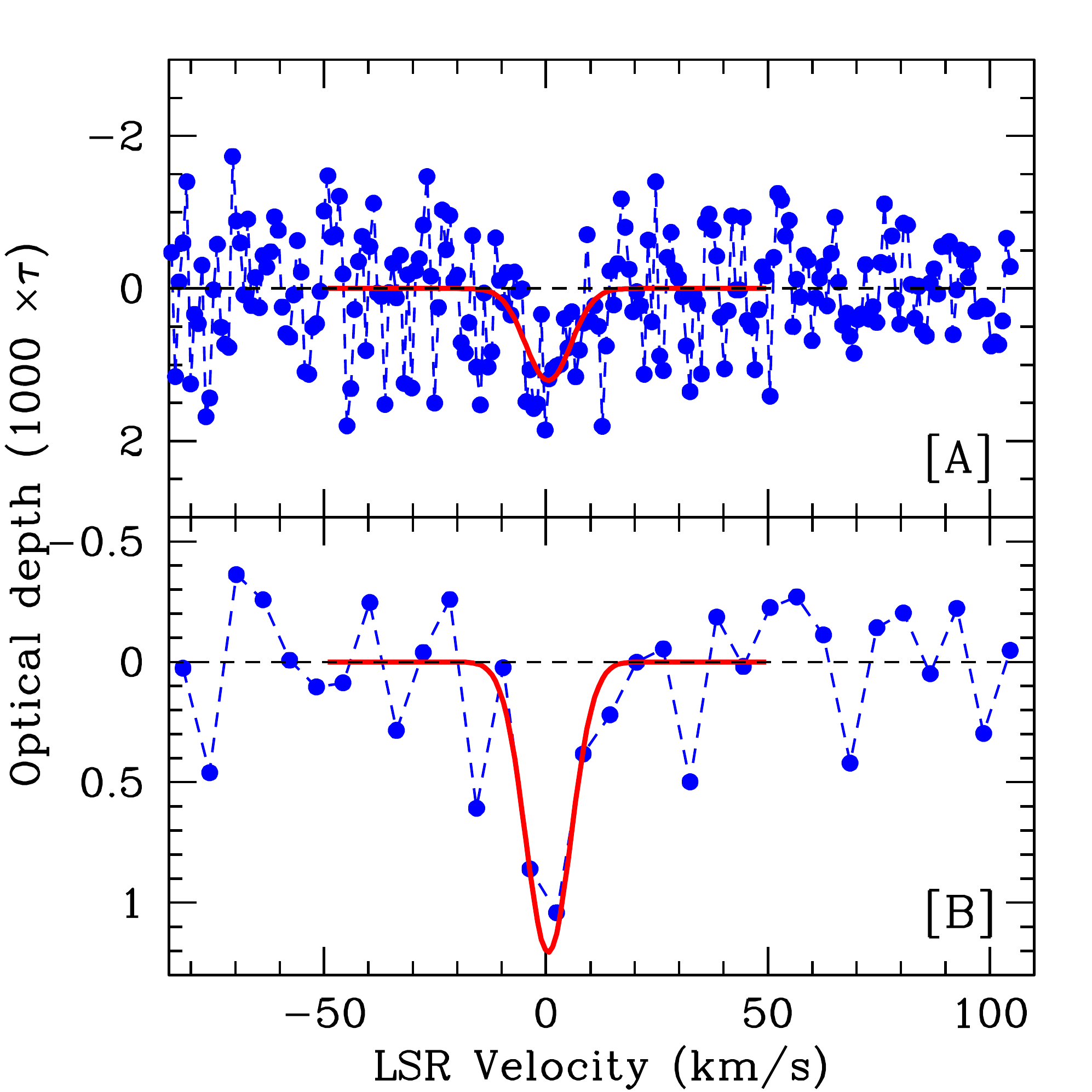}
\end{tabular}
\end{center}
\caption{The final GMRT \hii\ absorption spectrum towards B0438$-$436, at [A]~the Hanning-smoothed
and resampled velocity resolution of $\approx 0.86$~\kms\ (top panel), and 
[B]~smoothed to, and re-sampled at, a resolution of $\approx 6.0$~\kms\ (bottom panel).}
\label{fig:spc}
\end{figure}

\begin{figure}
\begin{center}
\begin{tabular}{c}
\includegraphics[width=3.3in]{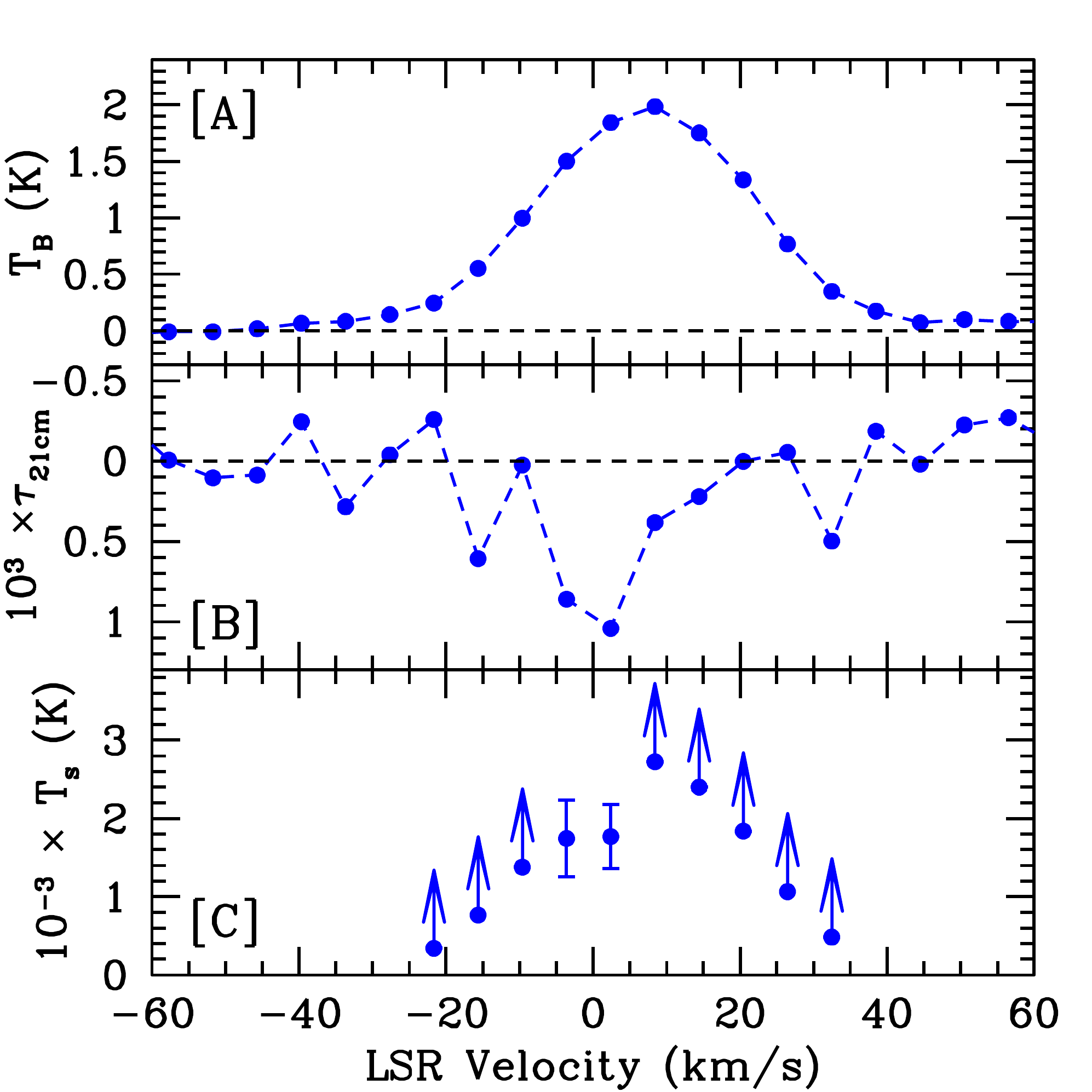}
\end{tabular}
\end{center}
\caption{The three panels of the figure show [A]~the \hii\ emission spectrum from the GASS survey 
	\citep{kalberla15}, [B]~the GMRT \hii\ optical depth spectrum, and [C]~the spin temperature 
	$\ts$, for the sightline towards B0438$-$436, plotted versus LSR velocity, in \kms.
	The spectra have been smoothed to, and re-sampled at, a velocity resolution of 
	$\approx 6.0$~\kms. See main text for discussion.}
	\label{fig:tspin}
\end{figure}

Neutral hydrogen is usually characterized by two ``temperatures'', the kinetic temperature, $\tk$,  and the 
\hii\ line excitation temperature, the spin temperature, $\ts$. In the case of the high-density CNM, the 
\hi\ hyperfine levels are expected to be thermalized by a combination of collisions and Lyman-$\alpha$ scattering, 
with the spin temperature approximately equal to the kinetic temperature \citep[e.g.][]{wouthuysen52,field58,liszt01}. 
Observationally, low spin temperatures ($\approx 100$~K) have indeed been found to be associated with the 
narrow \hii\ absorption features that are expected to arise in the CNM \citep[e.g.][]{dickey78,heiles03a,roy13a}. 
However, in the case of the WNM, the low number density implies that collisions are not very effective 
in thermalizing the \hii\ transition. Resonant scattering of Lyman-$\alpha$ photons is hence expected
to be the main process in driving $\ts$ towards $\tk$ \citep[e.g.][]{field65,deguchi85,liszt01}. It has long been 
unclear whether the fraction of the Galactic Lyman-$\alpha$ background radiation threading into the WNM is 
sufficient to drive the WNM spin temperature to the kinetic temperature \citep[e.g.][]{deguchi85,liszt01}.
Theoretical studies suggest that $\ts < \tk$ in the WNM, due to both the low WNM number density and the expected 
low penetration of background Lyman-$\alpha$ photons into the WNM \citep[e.g.][]{liszt01}. 
A few early observational studies did obtain high spin temperature estimates, $\gtrsim 5000$~K 
\citep[e.g.][]{kalberla80,payne83,kulkarni85}, albeit usually on sightlines with complex \hii\ absorption and 
with concerns about stray radiation affecting the \hii\ emission spectra. Further, while $\ts$ values 
$\gtrsim 1000$~K have recently been found at specific velocity ranges along multiple Galactic 
sightlines \citep[e.g.][]{carilli98b,dwaraka02,heiles03a,roy13a}, it is not usually straightforward to 
estimate $\tk$ at the same velocity ranges, making it difficult to test whether the 
\hii\ hyperfine excitation is indeed sub-thermal in the WNM.

Next, the \hi\ column density along a sightline is related to the velocity-integrated \hii\ optical depth 
by the equation 
\begin{equation}
\rm \nhi = 1.823 \times 10^{18} \langle T_s \rangle \int \tau_{21cm} dV \;,
\end{equation}
where $\rm \langle T_s \rangle$ is the column density-weighted harmonic mean spin temperature along the sightline.
Note that $\rm \langle T_s \rangle$ is biased towards low CNM temperatures: for example, if the \hi\ along the 
sightline is equally divided between phases with $\rm T_s \approx 100$~K and 
$\rm T_s \approx 8000$~K, one would obtain $\rm \langle T_s \rangle \approx 200$~K. Even if 90\% of the \hi\ along the 
sightline has $\rm T_s \approx 8000$~K, and 10\% has $\rm T_s \approx 100$~K, one would measure
$\rm \langle T_s \rangle \approx 900$~K. Exacerbating this issue, spin temperatures in the WNM are expected to be 
lower than the kinetic temperature, and are hence likely to be significantly lower than the assumed 
8000~K \citep[e.g.][]{liszt01}. As such, high values of $\rm  \langle T_s \rangle$, $\gtrsim 1000$~K, can only be 
obtained for sightlines with almost all the gas in the WNM. 

In the case of the sightline towards B0438$-$436, the GASS \hi\ emission spectrum yields an \hi\ 
column density of $1.29 \times 10^{20}$~\cm. Combining this with our measured integrated \hii\ optical 
depth of $\int \tau_{\rm 21cm} dV = (0.018 \pm 0.036)$~\kms\ then yields $\rm \langle T_s \rangle  = 3760 \pm 765$~K.
This is one of the highest column density-weighted harmonic mean spin temperatures ever measured along a 
sightline in the Milky Way, comparable to values seen in low-metallicity damped Lyman-$\alpha$
absorbers at high redshifts \citep[e.g.][]{kanekar03,kanekar14}. The sightline towards B0438$-$436
is clearly dominated by the WNM, with almost no cold gas present towards the quasar. The lack of 
CNM along this low-$\nhi$ sightline is consistent with the \hi\ column density threshold of 
$\approx 2 \times 10^{20}$~\cm\ that has been suggested for the formation of significant amounts 
of cold atomic gas in the Milky Way \citep{kanekar11b}.

Figs.~\ref{fig:tspin}[A] and [B] compare the GMRT \hii\ absorption spectrum towards 
B0438$-$436 with the GASS \hii\ emission spectrum at a neighbouring location, with both spectra
plotted versus LSR velocity, after smoothing to, and re-sampling at, a velocity resolution of 
6.0~\kms. It is clear that the peak \hii\ absorption occurs at a slightly lower velocity 
($\approx 0$~\kms) than the peak of the \hii\ emission ($\approx +8$~\kms). Indeed, the bulk of 
gas detected in \hii\ emission appears to not be detected in \hii\ absorption. 

Fig.~\ref{fig:tspin}[C] shows the spin temperature, $\rm T_s = T_B/(1 - e^{-\tau})$, plotted versus 
LSR velocity, at velocities where the \hii\ emission is detected at $\geq 5\sigma$ significance. 
Velocity channels with detections of \hii\ absorption at $\geq 3\sigma$ significance are shown as 
filled circles with error bars, while channels with non-detections of \hii\ absorption (i.e. with 
$< 3\sigma$ significance) are shown as $3\sigma$ lower limits to the spin temperature. While 
the two velocity channels with $\geq 3\sigma$ detections of absorption have $\rm T_s \approx 1750$~K,
the two higher-velocity channels have $\rm T_s(3\sigma) \gtrsim 2500$~K. These $\rm T_s$ 
values lie in the expected range for WNM spin temperatures \citep[e.g.][]{liszt01}.

The above results demonstrate that most or all of the gas towards B0438$-$436 is warm, with high spin 
temperatures, $\gtrsim 1750$~K, consistent with an origin in the stable WNM, at all channels with 
significant \hii\ emission. We also carried out a Gaussian decomposition of the \hii\ absorption profile 
of Fig.~\ref{fig:spc}[A], at a resolution of 0.86~\kms; a single Gaussian, shown as the solid curves 
in Figs.~\ref{fig:spc}[A] and [B], provides an excellent fit to the spectrum; this has a full width 
at half maximum (FWHM) of $12.0 \pm 3.1$~\kms, corresponding to a temperature of $3150 \pm 1635$~K. 
We emphasize that this 
estimate corresponds to the {\it maximum} allowed value of the kinetic temperature, sometimes referred 
to as the ``Doppler temperature'' \citep[$\rm T_D$; e.g.][]{roy13a}, as non-themal motions may contribute 
to the line broadening. This estimate of the Doppler temperature is very similar to the column 
density-weighted harmonic mean spin temperature estimate of $\rm \langle T_s \rangle = 3760 \pm 765$~K, 
and may suggest that the gas towards B0438$-$436 is in the unstable phase, with $\ts \approx \tk$. However, 
we emphasize that the above kinetic temperature estimate is consistent (within $\approx 1\sigma$ significance) 
with the standard stable WNM kinetic temperature range ($\approx 5000-8000$~K). This conclusion is 
rendered even more unlikely because $\ts \approx 1750$~K at the velocity channels with the strongest \hii\ 
absorption. Deeper \hii\ absorption studies are needed to accurately estimate $\tk$ via the Gaussian 
decomposition approach.

\begin{figure*}
\begin{center}
\includegraphics[width=3.3in]{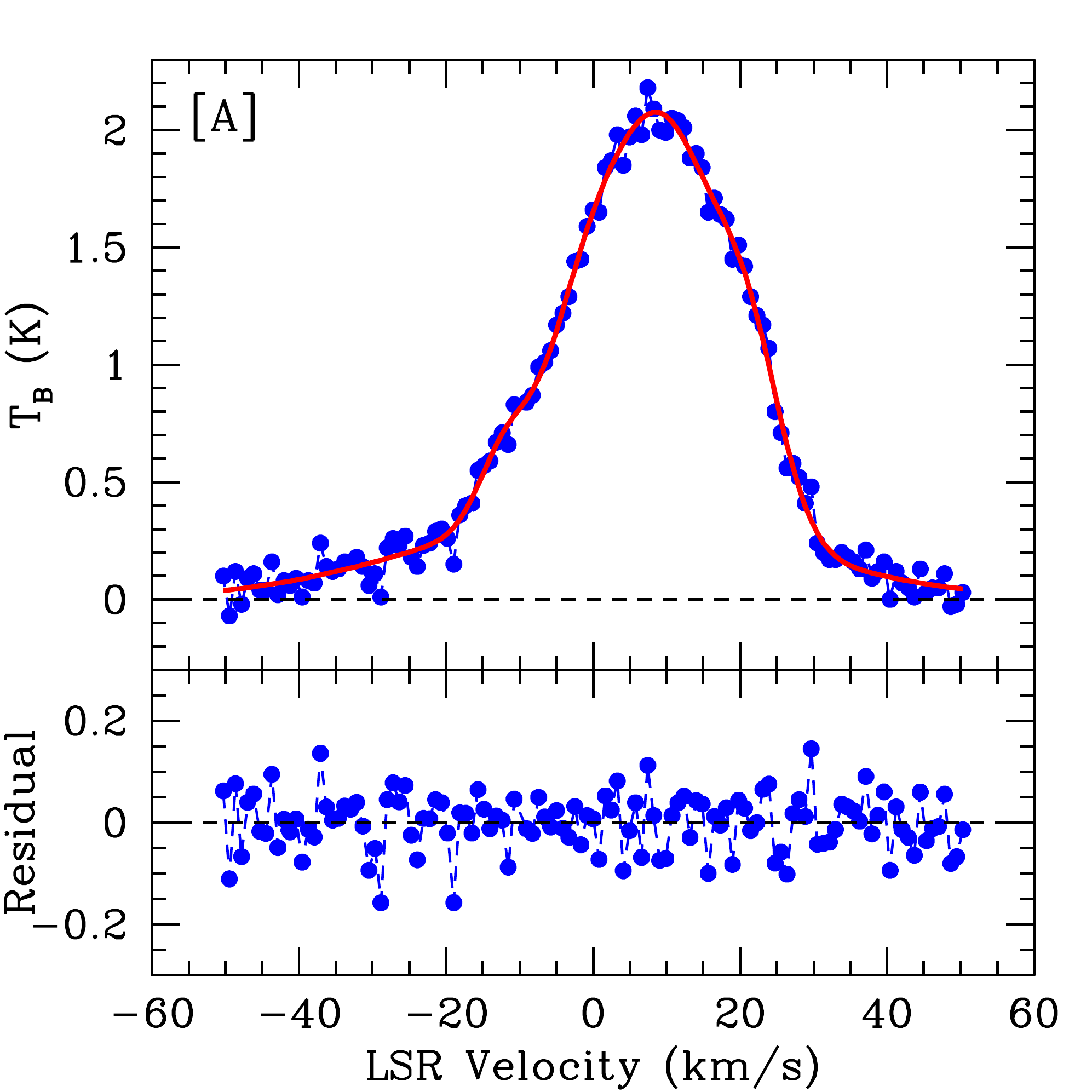}
\includegraphics[width=3.3in]{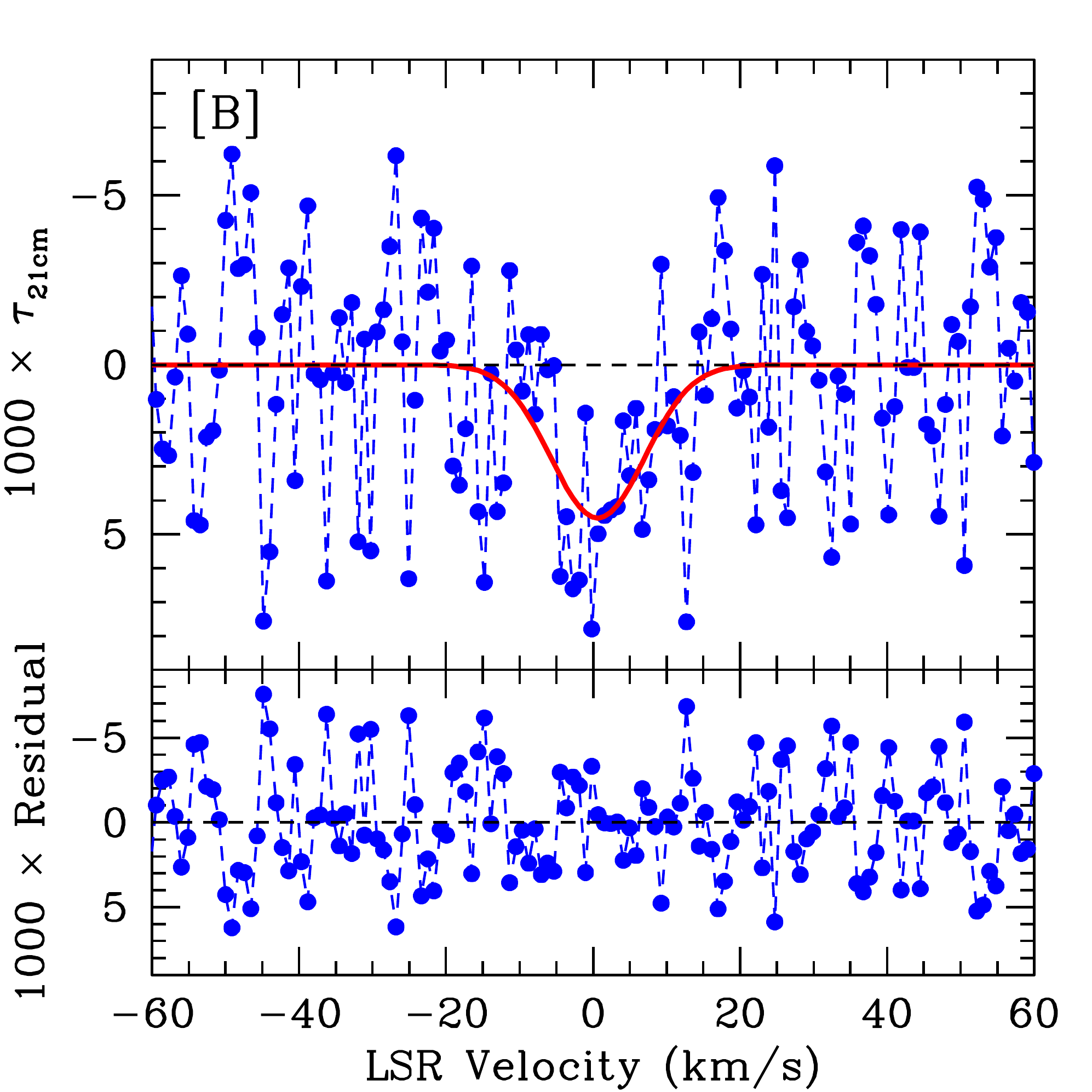}
\end{center}
\caption{Results of the multi-Gaussian joint decomposition of the [A]~\hii\ emission and [B]~\hii\
	absorption spectra; the former spectrum is at a resolution of $\approx 0.82$~\kms, while the 
	latter is at a resolution of $\approx 0.86$~\kms. The top panels show the best-fit model 
	(solid curve) overlaid on the two spectra, while the bottom two panels show the residuals 
	from the fit.}
\label{fig:gauss}
\end{figure*}

\setcounter{table}{0}
\begin{table*}
\begin{center}
\caption{Gaussian joint decomposition of the \hii\ emission and absorption spectra.
\label{tab:gauss}}
\begin{tabular}{|c|c|c|c|c|c|c|}
\hline 
Component & $\rm T_B$$^a$ & V$_0$$^b$ & FWHM  & $\tau_{\rm pk}$$^c$ & $\tk$$^d$ &  $\ts$$^e$ \\
	  &    K          &   \kms\   & \kms\ & $\times 10^{-3}$    &  K        &  K  \\
\hline                                                                                                                                       
1 & $1.08 \pm 0.30$   & $0.63 \pm 1.57$ & $15.0 \pm 2.9$ & $1.08 \pm 0.22$ & $4910 \pm 1900$   & $1000 \pm 345$ \\
2 & $1.31 \pm 0.35$   & $10.9 \pm 1.9$  & $13.6 \pm 4.8$ & $< 0.48$        & $4045 \pm 2850$   & $> 2730$ \\
3 & $0.358 \pm 0.067$ & $0.90 \pm 1.46$ & $56.9 \pm 5.0$ & $< 0.23$        & $70615 \pm 12410$ & $> 1555$ \\
4 & $0.26 \pm 0.10$   & $-12.1 \pm 1.1$ & $8.4 \pm 2.2$  & $< 0.61$        & $1540 \pm 805$    & $> 425$ \\
5 & $0.81 \pm 0.52$   & $20.8 \pm 2.4$  & $11.4 \pm 2.1$ & $< 0.52$        & $2835 \pm 1045$   & $> 1560$ \\
\hline
\end{tabular}
\end{center}
\begin{center}
For each component, the columns are: $^a$The peak brightness temperature; $^b$The central LSR 
velocity; $^c$The peak \hii\ optical depth, or, for non-detections, the $3\sigma$ upper limit on the 
peak \hii\ optical depth, at a velocity resolution equal to the line FWHM; $^d$The kinetic 
temperature inferred from the FWHM; $^e$The spin temperature or, for components 2--5, the 
$3\sigma$ lower limit to the spin temperature. \\
\end{center}
\end{table*}

Attempts have been made in the literature to carry out a joint multi-Gaussian decomposition 
of the \hii\ emission and the \hii\ absorption spectra \citep[e.g.][]{heiles03a,murray14}. These 
have usually encountered difficulties in handling \hii\ ``self-absorption'' along the sightline, by 
foreground CNM clouds of their own, and of background CNM and WNM, emission. The present sightline
towards B0438$-$436 is very interesting in this context because the low peak \hii\ optical
depth ($\approx 0.001$) means that self-absorption is not an issue. The sightline thus allows 
the possibility of a joint Gaussian decomposition of the \hii\ absorption and emission spectra 
without any effects of self-absorption. 

We carried out a joint multi-Gaussian fit to the \hii\ emission and \hii\ absorption spectra, the former
at the original velocity resolution of $\approx 0.82$~\kms, and the latter after Hanning-smoothing
and resampling, at a resolution of $\approx 0.86$~\kms. We allowed  for \hii\ absorption by a single Gaussian 
component and \hii\ emission by multiple Gaussian components. In the fit, the position and FWHM of one 
of the emission components was constrained to be the same as that of the absorption component (using
multiple absorption components did not yield a better fit, in terms of an improved reduced chi-square 
value). The best-fit model, with a reduced chi-square $\chi^2_N = 1.00$, contains 5 emission components 
and one absorption component. The model is plotted in Figs.~\ref{fig:gauss}[A]
and [B], with the residuals from the fit displayed in the lower two panels; the residuals are seen to be 
consistent with noise. The properties of the different Gaussian components are summarized in 
Table~\ref{tab:gauss}, whose columns contain (1)~the component number, (2)~the peak brightness 
temperature of the component, in K, (3)~the central LSR velocity, in \kms, (4)~the component FWHM, 
in \kms, (5)~the peak \hii\ optical depth, or the $3\sigma$ upper limit to the peak optical depth 
(estimated at a velocity resolution equal to the component FWHM), (6)~the kinetic temperature of the 
component, $\rm T_k$, and (7)~the component spin temperature $\rm T_s$, or the $3\sigma$ 
lower limit to the spin temperature.

Initially, we note that the third component in Table~\ref{tab:gauss} is extremely wide, with an FWHM of 
$\approx 57$~\kms. Given the relatively high Galactic latitude of B0438$-$436 ($b \approx -41^\circ$), it 
appears unlikely that this large broadening arises from either Galactic rotation or turbulence. We suspect 
that it may be due to a residual spectral baseline in the GASS spectrum, although we note that a similar 
wide component is also visible in the \hii\ spectrum from the Leiden-Argentine-Bonn survey \citep{kalberla05}. 
For now, the origin of the broad component remains unclear.

The component with detected \hii\ absorption has an FWHM of $15.0 \pm 2.9$~\kms, in good agreement with
that obtained from the fit to the \hii\ absorption alone ($12.0 \pm 3.1$~\kms). The inferred upper limit on the 
kinetic temperature is then $4910 \pm 1900$~K, in the stable WNM temperature range. We emphasize, as above, 
that this is formally an estimate of the Doppler temperature $\td$, and hence only yields an upper limit to 
$\tk$, due to the possibility of non-thermal broadening. 

We consider the two extreme possibilities for line broadening: (1) weak non-thermal motions, for 
which the kinetic temperature is approximately equal to the Doppler temperature, i.e. $\tk \approx \td = 
(4910 \pm 1900)$~K, and (2)~strong non-thermal broadening, in which case $\tk \ll \td$. In the first case, 
the detected \hii\ absorption arises from the stable WNM. The spin temperature of this component is $1000 \pm 345$~K, 
significantly lower than the kinetic temperature. Figs.~2 and 5 of \citet{liszt01} suggest that this would 
require the gas to be at low pressures, $P/k \lesssim 1000$~cm$^{-3}$~K, and with almost no Galactic 
Lyman-$\alpha$ background penetrating into the \hi. The data would then support a picture 
of little coupling between the WNM and the background Lyman-$\alpha$ radiation field.

In the second case, of strong non-thermal broadening, the measured Doppler temperature only yields an upper
limit to $\tk$. However, the measured spin temperature gives the constraint $\tk \geq 1000$, 
since $\tk \geq \ts$ in the WNM. This implies the range $\rm 1000 \: K \: \leq \tk \leq 4910\: K$ for 
the component detected in absorption. For strong non-thermal broadening, $\tk \ll \td$, the detected 
\hii\ absorption would then still arise in warm gas, albeit from the thermally unstable phase, 
with $\tk \approx \ts$, i.e.  with strong coupling between the unstable neutral medium and the 
background Lyman-$\alpha$ radiation field. We cannot rule out this possibility with the present data.

Finally, the Doppler temperature of component~2 of Table~\ref{tab:gauss} is marginally consistent with its
origin in the stable WNM, albeit with large errors. It is interesting that this component has a relatively 
high lower limit on its spin temperature, $\ts > 2730$~K, which suggests a stronger coupling with the 
background Lyman-$\alpha$ radiation. The Doppler temperatures of components~4 and 5 appear consistent with 
an origin in the unstable phase, although again with large errors. Again, it is interesting that the 
lower limit on the spin temperature of component~5 is relatively high, approaching the Doppler temperature. 
We emphasize, however, that the errors in the Doppler temperature estimates are still quite large. Deeper 
observations of the \hii\ absorption towards B0438$-$36 are needed to confirm the above results, both 
the absolute values of the Doppler temperatures, and hence, the relation between the spin and the kinetic 
temperatures in the stable WNM and the unstable neutral medium.

\section{Summary}

We have used the GMRT to detect wide, weak Galactic \hii\ absorption towards the quasar B0438$-$436.
The velocity-integrated \hii\ optical depth is very low, $\int \tau d{\rm V} = (0.0188 \pm 0.0036)$~\kms. Combining 
this with the \hi\ column density ($\nhi = 1.29 \times 10^{20}$~\cm) measured from the GASS \hii\ emission spectrum on 
a neighbouring sightline yields a column density-weighted harmonic mean spin temperature of $\rm \langle T_s \rangle = (3760 \pm 765)$~K, 
far higher than typical in the Galaxy. The high $\rm \langle T_s \rangle$ value indicates that there is little or no CNM along the 
sightline towards B0438$-$436. A detailed comparison between the \hii\ emission and absorption spectra indicates that
$\ts \geq 1760$~K over the range of velocities showing the strongest \hii\ emission, consistent with the spin temperature 
range expected in the stable WNM. The extremely low \hii\ optical depth further implies that \hii\ self-absorption 
is negligible along this sightline. We carry out a joint multi-Gaussian decomposition of the \hii\ emission and the 
\hii\ absorption profiles, to obtain $\tk \leq (4910 \pm 1900)$~K for the component with detected 
\hii\ absorption, consistent with the stable WNM temperature range. The spin temperature of this 
component is $(1000 \pm 345)$~K, significantly lower than the kinetic temperature. This suggests that the 
\hii\ line excitation is sub-thermal in the WNM along this sightline, possibly due to low gas pressure or 
low threading of the Galactic Lyman-$\alpha$ background into the WNM. However, the present data cannot rule 
out the possibility that the detected \hii\ absorption might arise from warm gas in the thermally-unstable phase, with
significant non-thermal line broadening, and the gas spin temperature comparable to the kinetic temperature.

\section*{Acknowledgments}

We thank the GMRT staff who have made these observations possible. The GMRT is run by the 
National Centre for Radio Astrophysics of the Tata Institute of Fundamental Research.
NK acknowledges support from the Department of Science and Technology via 
a Swarnajayanti Fellowship (DST/SJF/PSA-01/2012-13). NR acknowledges support from 
the Infosys Foundation through the Infosys Young Investigator grant.
\bibliographystyle{mn2e}
\bibliography{ms}

\end{document}